\documentclass[prb,twocolumn,showkeys,showpacs,amsmath,amsfonts,floatfix,superscriptaddress,nofootinbib]{revtex4-1}

\usepackage{amssymb,amsmath,amstext}                
\usepackage{graphicx}                                               
\usepackage{epstopdf}                                               
\usepackage{color}       
\usepackage{bm}
\usepackage{appendix}                                              
\usepackage[latin2]{inputenc}
\usepackage{ulem}
\usepackage{latexsym}
\usepackage[colorlinks=true,citecolor=blue,linkcolor=magenta]{hyperref}
\def\be{\begin{equation}}
\def\ee{\end{equation}}
\def\bea{\begin{eqnarray}}
\def\eea{\end{eqnarray}}
\def\bi{\begin{itemize}}
\def\ei{\end{itemize}}

\begin{document}

\title{ Time evolution of an infinite projected entangled pair state: \\
                     a neighborhood tensor update  }

\author{Jacek Dziarmaga}
\email{dziarmaga@th.if.uj.edu.pl}
\affiliation{Jagiellonian University, Institute of Theoretical Physics, 
             ul. \L{}ojasiewicza 11, 30-348 Krak\'ow, Poland }

\date{\today}

\begin{abstract}
The simple update (SU) and full update (FU) are the two paradigmatic time evolution algorithms for a tensor network known as the infinite projected entangled pair state (iPEPS). They differ by an error measure that is either, respectively, local or takes into account full infinite tensor environment. In this paper we test an intermediate neighborhood tensor update (NTU) accounting for the nearest neighbor environment.
This small environment can be contracted exactly in a parallelizable way. 
It provides an error measure that is Hermitian and non-negative down to machine precision.
In the 2D quantum Ising model NTU is shown to yield stable unitary time evolution following a sudden quench.
It also yields accurate thermal states despite correlation lengths that reach up to 20 lattice sites.
The latter simulations were performed with a manifestly Hermitian purification of a thermal state. 
Both were performed with reduced tensors that do not include physical (and ancilla) indices.
This modification naturally leads to two other schemes: 
a local SVD update (SVDU) and a full tensor update (FTU) being a variant of FU.
\end{abstract}

\maketitle


\section{Introduction}
\label{sec:introduction}

Weakly entangled states are just a small subset in an exponentially large Hilbert space but they are ubiquitous as stationary ground or thermal states in condensed matter physics. They can be efficiently represented by tensor networks~\cite{Verstraete_review_08,Orus_review_14}, including the one-dimensional (1D) matrix product state (MPS)~\cite{fannes1992}, its two-dimensional (2D) generalization known as a projected entangled pair state (PEPS)~\cite{Nishino_2DvarTN_04,verstraete2004}, or a multi-scale entanglement renormalization ansatz~\cite{Vidal_MERA_07,Vidal_MERA_08,Evenbly_branchMERA_14,Evenbly_branchMERAarea_14}. The MPS ansatz provides a compact representation of ground states of 1D gapped local Hamiltonians~\cite{Verstraete_review_08,Hastings_GSarealaw_07,Schuch_MPSapprox_08} and purifications of their thermal states~\cite{Barthel_1DTMPSapprox_17}. It is also the ansatz underlying the density matrix renormalization group (DMRG)~\cite{White_DMRG_92, White_DMRG_93,Schollwock_review_05,Schollwock_review_11}. Analogously, the 2D PEPS is expected to represent ground states of 2D gapped local Hamiltonians~\cite{Verstraete_review_08,Orus_review_14} and their thermal states~\cite{Wolf_Tarealaw_08,Molnar_TPEPSapprox_15}, though representability of area-law states, in general, was shown to have its limitations~\cite{Eisert_TNapprox_16}. As a variational ansatz tensor networks do not suffer from the notorious sign problem plaguing the quantum Monte Carlo methods. Consequently, they can deal with fermionic systems~\cite{Corboz_fMERA_10,Eisert_fMERA_09,Corboz_fMERA_09,Barthel_fTN_09,Gu_fTN_10}, as was shown for both finite~\cite{Cirac_fPEPS_10} and infinite PEPS (iPEPS)~\cite{Corboz_fiPEPS_10,Corboz_stripes_11}.

\begin{figure}[t!]
\vspace{-0cm}
\includegraphics[width=0.9999\columnwidth,clip=true]{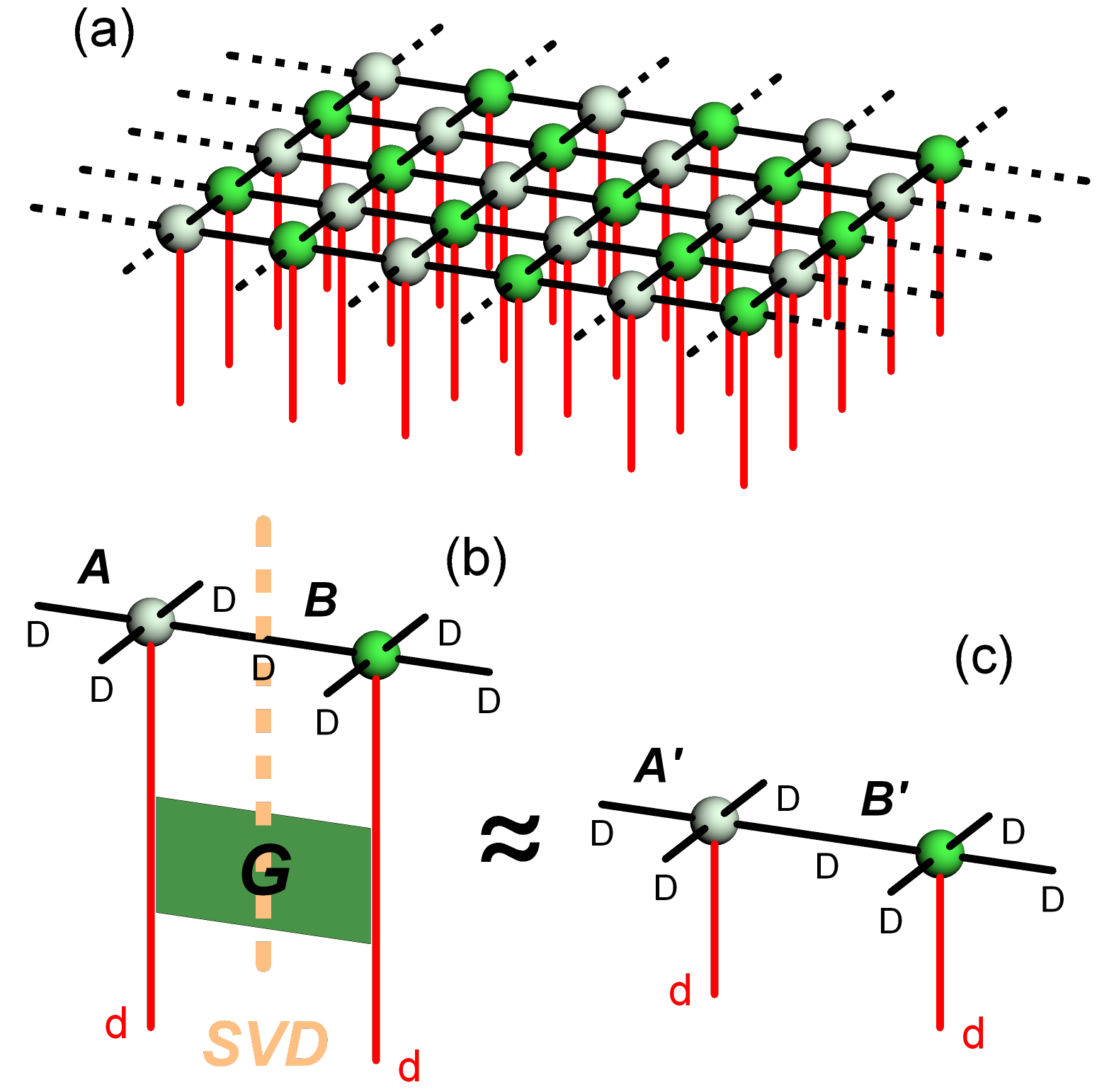}
\vspace{-0cm}
\caption{
{\bf Schematic SVD update (SVDU). }
In (a)
an infinite square lattice is divided into two sublattices with tensors $A$ (lighter) and $B$ (darker).
Their contraction is the iPEPS tensor network.
The red lines are physical indices.
In (b)
a two-site Suzuki-Trotter gate $G$ is applied to physical indices of every considered NN pair of tensors $A$ and $B$. Then the contraction is subject to a singular value decomposition (SVD) between left and right indices separated by the dashed orange line. The SVD of this $dD^3\times dD^3$ matrix, $U\lambda V^\dag$, is truncated to $D$ leading singular values.
In (c) 
after the truncation new tensors are obtained as $A'=U\sqrt{\lambda}$ and $\sqrt{\lambda}V^\dag=B'$.
This is just a schematic form of SVDU. 
Its efficient implementation is explained in Fig. \ref{fig:2site2} together with its upgrade to the neighborhood update (NTU) and full tensor update (FTU).
}
\label{fig:2site}
\end{figure}

The PEPS was originally proposed as an ansatz for ground states of finite systems~\cite{Verstraete_PEPS_04, Murg_finitePEPS_07}, generalizing earlier attempts to construct trial wave-functions for specific models~\cite{Nishino_2DvarTN_04}. The subsequent development of efficient numerical methods for infinite PEPS (iPEPS)~\cite{Cirac_iPEPS_08,Xiang_SU_08,Gu_TERG_08,Orus_CTM_09}, shown in Fig. \ref{fig:2site}(a), promoted it as one of the methods of choice for strongly correlated systems in 2D. Its power was demonstrated, e.g., by a solution of the long-standing magnetization plateaus problem in the highly frustrated compound $\textrm{SrCu}_2(\textrm{BO}_3)_2$~\cite{matsuda13,corboz14_shastry}, establishing the striped nature of the ground state of the doped 2D Hubbard model~\cite{Simons_Hubb_17}, and new evidence supporting gapless spin liquid in the kagome Heisenberg antiferromagnet~\cite{Xinag_kagome_17}. Recent developments in iPEPS optimization~\cite{fu,Corboz_varopt_16,Vanderstraeten_varopt_16}, contraction~\cite{Fishman_FPCTM_17,Xie_PEPScontr_17}, energy extrapolations~\cite{Corboz_Eextrap_16}, and universality-class estimation~\cite{Corboz_FCLS_18,Rader_FCLS_18,Rams_xiD_18} pave the way towards even more complicated problems, including simulation of thermal states~\cite{Czarnik_evproj_12,Czarnik_fevproj_14,Czarnik_SCevproj_15, Czarnik_compass_16,Czarnik_VTNR_15,Czarnik_fVTNR_16,Czarnik_eg_17,Dai_fidelity_17,CzarnikDziarmagaCorboz,czarnik19b,Orus_SUfiniteT_18,CzarnikKH,wietek19,jimenez20,poilblanc20,CzarnikSS}, mixed states of open systems~\cite{Kshetrimayum_diss_17,CzarnikDziarmagaCorboz}, excited states~\cite{Vanderstraeten_tangentPEPS_15,ExcitationCorboz}, or real-time evolution~\cite{CzarnikDziarmagaCorboz,HubigCirac,tJholeHubig,Abendschein08,SUlocalization,SUtimecrystal}. In parallel with iPEPS, there is continuous progress in simulating systems on cylinders of finite width using DMRG. This numerically highly stable method that is now routinely used to investigate 2D ground states~\cite{Simons_Hubb_17,CincioVidal} was applied also to thermal states on a cylinder~\cite{Stoudenmire_2DMETTS_17,Weichselbaum_Tdec_18,WeichselbaumTriangular,WeichselbaumBenchmark,chen20}. However, the exponential growth of the bond dimension limits the cylinder's width to a few lattice sites. Among alternative approaches are direct contraction and renormalization of a 3D tensor network representing a 2D thermal density matrix \cite{Li_LTRG_11,Xie_HOSRG_12,Ran_ODTNS_12,Ran_NCD_13,Ran_THAFstar_18,Su_THAFoctakagome_17,Su_THAFkagome_17,Ran_Tembedding_18}.

This article readdresses the problem of real/imaginary time evolution with iPEPS\cite{CzarnikDziarmagaCorboz}. There are two most popular simulation schemes: the simple update (SU) and full update (FU). In both the time evolution proceeds by small time steps, each of them subject to the Suzuki-Trotter decomposition. In both after a Trotter gate is applied to a pair of nearest neighbor (NN) sites a bond dimension of the index between the sites is increased by a factor equal to the rank of the gate. In order to prevent its exponential growth with time the dimension is truncated to a predefined value, $D$, in a way that minimizes error incurred by the truncation. The two schemes differ by a measure of the error: FU takes into account full infinite tensor environment while SU only the bonds adjacent to the NN sites. The former is expected to perform better in case of long range correlations while the latter is, at least formally, more efficient thanks to its locality. In this paper an intermediate scheme is considered --- a neighborhood tensor update (NTU) --- where the error measure is induced by the sites that are NN to the Trotter gate. It is shown to compromise the FU accuracy only a little for a price of small numerical overhead over SU, hence it may turn out to be a reasonable trade off for many applications.

The neighborhood tensor update is a special case of a cluster update \cite{wang2011cluster} where the size of the environment is a variable parameter interpolating between a local update and the infinite FU. In NTU only the neighboring sites are taken into account because they allow the error measure to be calculated exactly with little numerical overhead over SU as it involves only tensor contractions that are fully parallelizable. Its exactness warrants the error measure to be a manifestly Hermitian and non-negative quadratic form. This property is essential for stability of NTU and makes it distinct from FU where an approximate corner transfer matrix renormalization \cite{Orus_CTM_09,Orus_review_14} often breaks the Hermiticity and non-negativeness. In case of long range correlations the small environment can, admittedly,  make NTU converge with the bond dimension more slowly than FU but this may be compensated by its better numerical efficiency and stability that allow NTU to reach higher bond dimensions.

At a more technical level, unlike in FU but similarly as in Ref. \onlinecite{Evenbly2018}, we define reduced tensors not before but after application of the Trotter gate. Our reduced tensors do not have any physical (and ancilla) indices. Unlike in Ref. \onlinecite{Evenbly2018}, we do not introduce any bond tensors in our iPEPS to avoid the necessity of their inversion. This redefinition of the reduced tensors naturally leads to two schemes that are complementary to NTU: a local SVD update (SVDU) and a full tensor update (FTU). The former is more local than SU, as it ignores even the adjacent bonds' environment, while the latter is a variant of FU with the infinite but approximate environment. 

Another technical modification, in case of thermal states represented by their purifications, is to make the purification manifestly Hermitian between physical and ancilla degrees of freedom. The Hermitian purification is an iPEPS in a space of Hermitian operators. An important symmetry is protected thus enhancing stability and in general also numerical efficiency.

This paper is organized as follows. In section \ref{sec:algorithms} we provide a detailed introduction to SVDU, NTU, and FTU that includes the definition of reduced tensors. In section \ref{sec:sudden} the algorithms are applied to unitary real time evolution after a sudden quench of the 2D quantum Ising Hamiltonian. In section \ref{sec:thermal} we describe the manifestly Hermitian thermal state purifications and in section \ref{sec:thermalIsing} thermal states of the 2D quantum Ising model are simulated by imaginary time evolution of their purifications. We summarize in section \ref{sec:conclusion}. 


\begin{figure}[t!]
\vspace{-0cm}
\includegraphics[width=0.9999\columnwidth,clip=true]{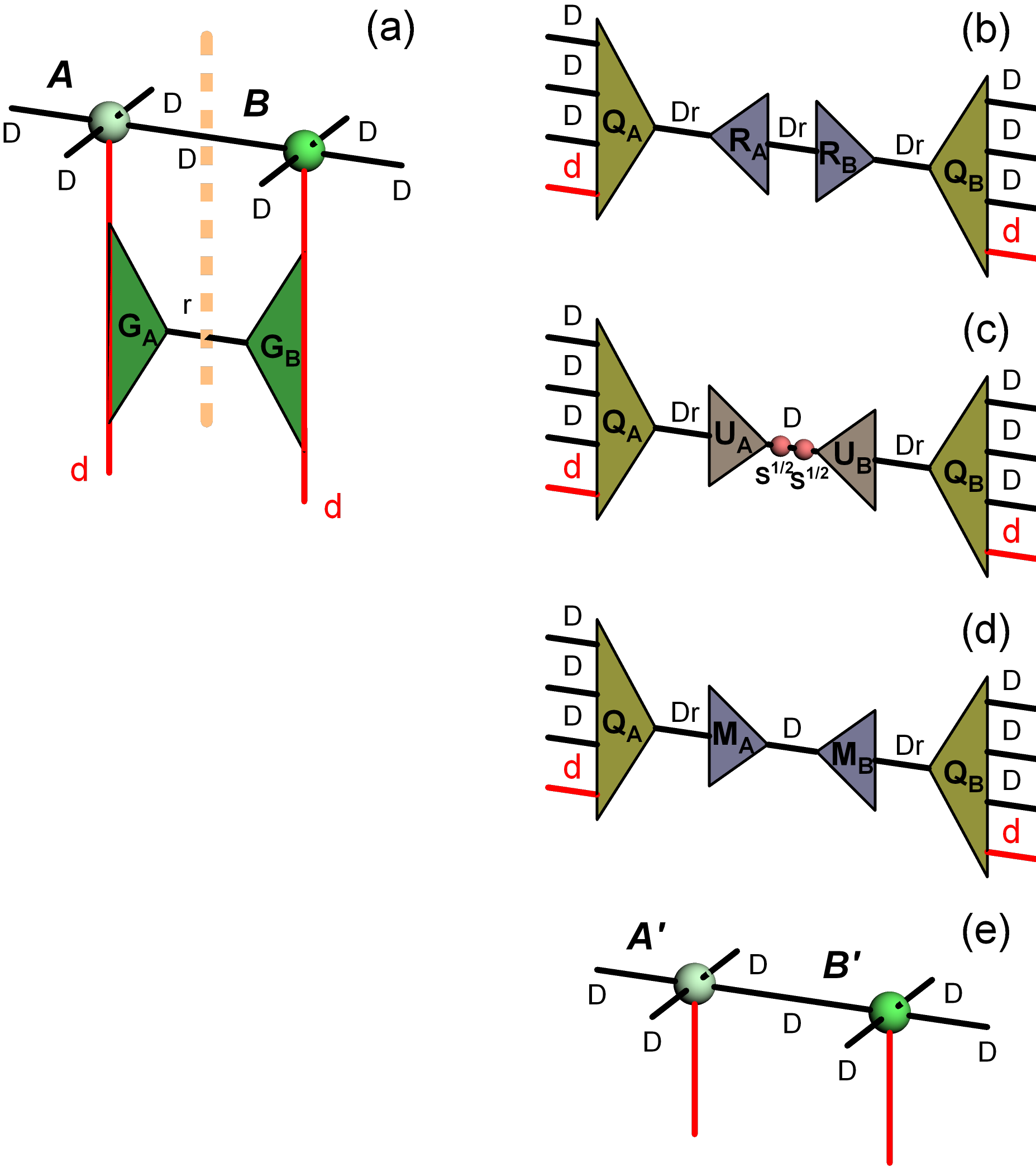}
\vspace{-0cm}
\caption{
{\bf Efficient SVDU and beyond. }
In (a) 
the Suzuki-Trotter gate $G$ is applied to physical indices of NN tensors $A$ and $B$ as in Fig. \ref{fig:2site}(b). Here the 2-site gate is replaced by two tensors, $G_A$ and $G_B$, contracted by an index with dimension $r$.
In (b)
the tensor contraction $A\cdot G_A$ is QR-decomposed into $Q_AR_A$. Similarly $B\cdot G_B=Q_BR_B$. 
Isometries $Q_{A,B}$ will remain fixed.
In (c)
after SVD, $R_AR_B^T=U_ASU_B^T$, $S$ is truncated to $D$ leading singular values.
In (d)
matrices $M_A=U_AS^{1/2}$ and $M_B^T=S^{1/2}U_B^T$ are made by absorbing square root of truncated $S$ symmetrically.
In (e)
in SVDU new iPEPS tensors are obtained as $A'=Q_A\cdot M_A$ and $B'=Q_B\cdot M_B$ ending the story.
In the FTU/NTU schemes matrices $M_{A,B}$ are optimized in full/neighborhood tensor environment, see Fig. \ref{fig:env}, before being contracted with isometries $Q_{A,B}$ to make $A'$ and $B'$.
}
\label{fig:2site2}
\end{figure}

\begin{figure}[t!]
\vspace{-0cm}
\includegraphics[width=0.9999\columnwidth,clip=true]{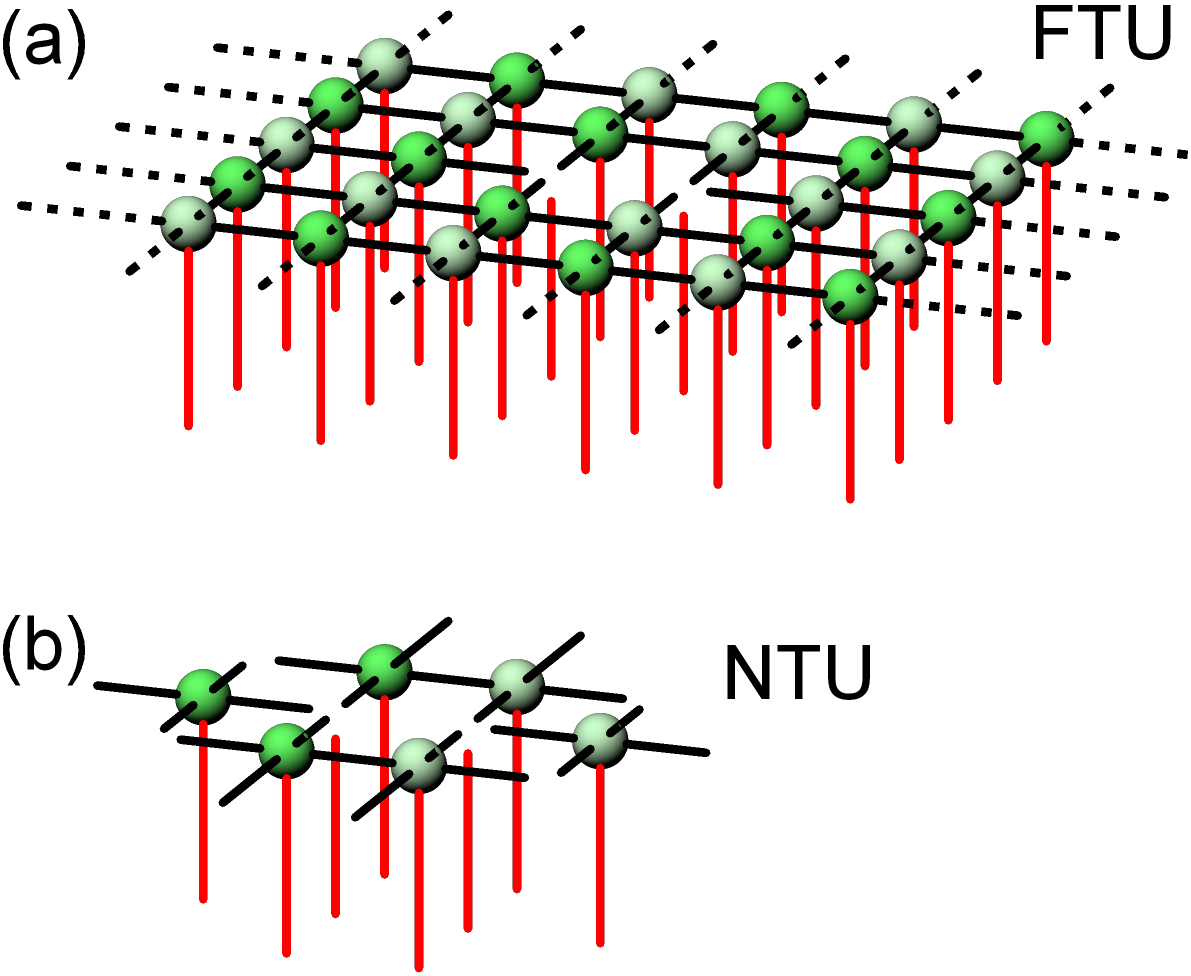}
\vspace{-0cm}
\caption{
{\bf Full/neighborhood tensor update. }
In FTU the truncated diagram in Fig. \ref{fig:2site2}(d) is inserted in place of the two missing tensors in the middle of the infinite PEPS in (a). 
In NTU the same truncated diagram is inserted into a finite fragment of the iPEPS in (b). 
In FTU/NTU the tensor network obtained after the truncated insertion is compared with a similar network (a)/(b) but inserted with the exact diagram in Fig. \ref{fig:2site2}(b). 
In both schemes truncated matrices $M_A$ and $M_B$ are optimized to minimize norm of the difference between the two networks: the one with the truncated insertion and the other with the exact one.
}
\label{fig:env}
\end{figure}

\begin{figure}[t!]
\vspace{-0cm}
\includegraphics[width=0.70\columnwidth,clip=true]{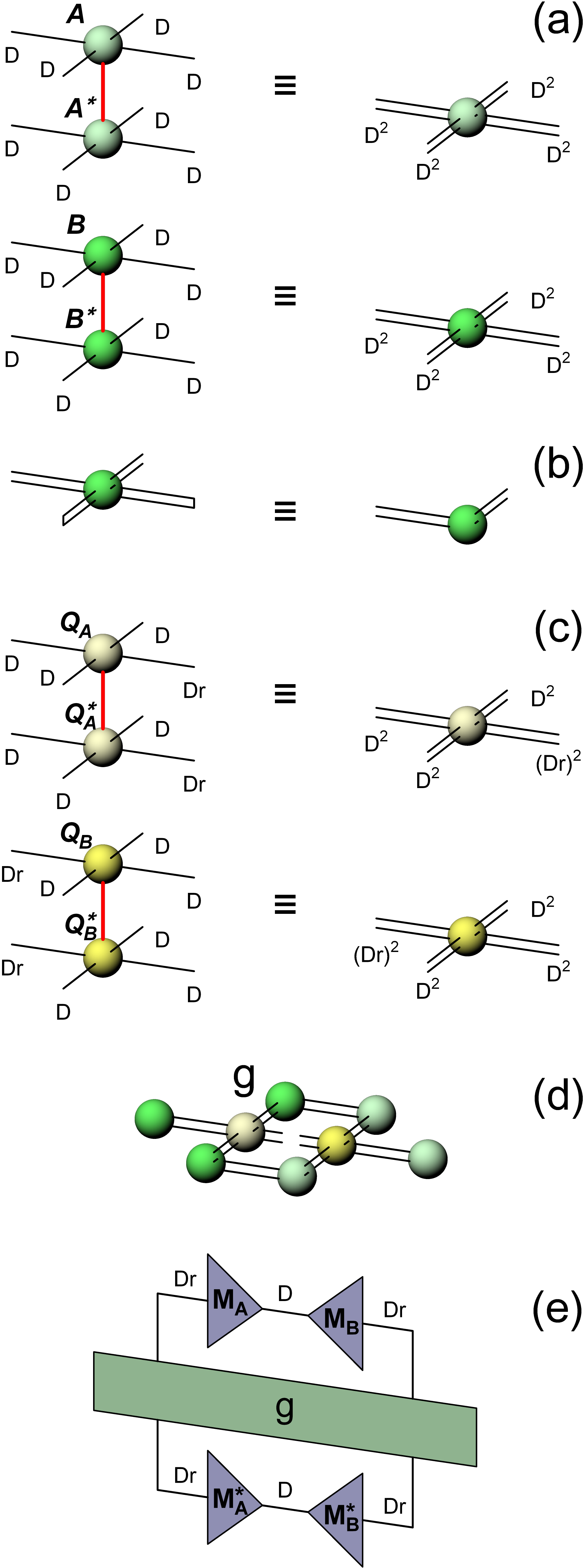}
\vspace{-0cm}
\caption{
{\bf NTU. }
Norm squared of the matrix product,
$\vert\vert M_AM_B^T \vert\vert^2$, is shown in (e).
Here $g$ is a metric tensor obtained in (d). The upper/lower free indices in diagram (d) correspond to the upper/lower indices of $g$ in (e).
Diagram (d) is obtained by inserting the two empty central sites in Fig. \ref{fig:env}(b) with isometries $Q_A$ and $Q_B$ and then contracting the inserted (ket) network with its complex conjugate (bra) trough pairs of their corresponding indices, except for the bond indices stemming from the two isometries along the considered bond. This is done is steps (a,b,c,d).
In (a) double iPEPS tensors are defined. Edge double tensors are obtained by contracting pairs of corresponding external bra and ket indices, as exemplified in (b). In (c) double isometries are defined. The double isometries and edge tensors are assembled in (d). Cost of their contractions can be optimized to scale like $D^8$.  
}
\label{fig:NTUenv}
\end{figure}

\begin{figure}[t!]
\vspace{-0cm}
\includegraphics[width=0.5\columnwidth,clip=true]{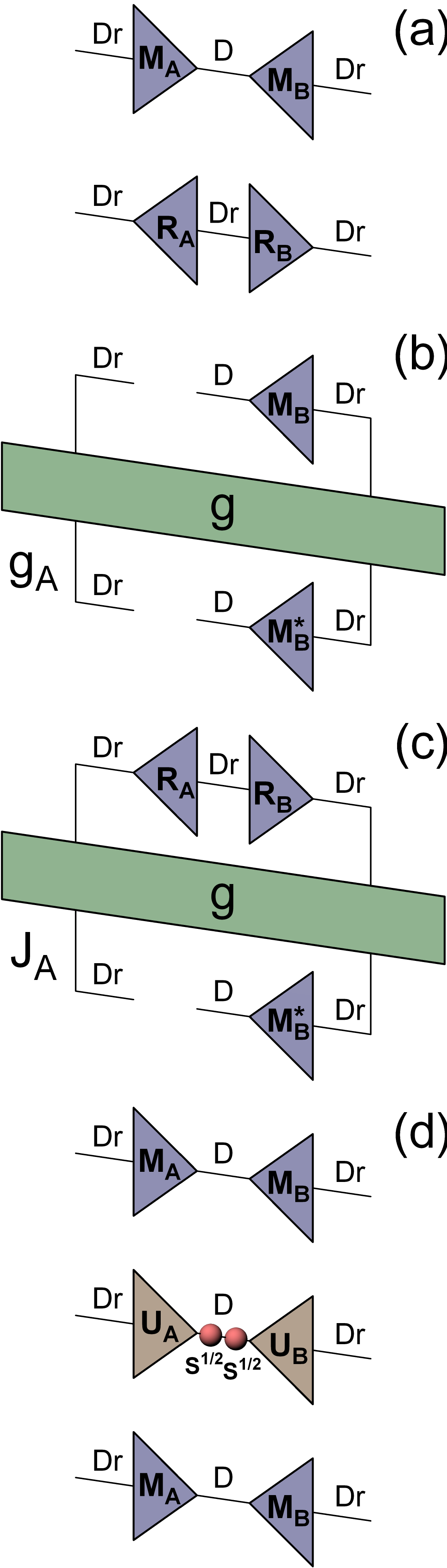}
\vspace{-0cm}
\caption{
{\bf Optimization in NTU/FTU. }
In (a)
matrices $M_A, M_B$ are optimized for their product, $M_A M_B^T$, to be the best approximation to the exact product, $R_AR_B^T$. The error is measured with the metric in Fig. \ref{fig:NTUenv}(e). 
In (b)
reduced metric tensor $g_A$ for matrix $M_A$.
In (c)
reduced source term $J_A$ for matrix $M_A$.
In (d) 
a product of converged matrices is subject to a SVD, $M_A M_B=U_ASU_B^T$, after which new balanced matrices, $M_A=U_AS^{1/2}$ and $M_B^T=S^{1/2}U_B^T$, are formed by absorbing singular values $S$ in a symmetric way. 
However, iterative optimization of the matrices is not symmetric. Before optimization with respect to $M_A$ the matrices are ``tilted'' as $M_A=U_AS$ and $M_B^T=U_B^T$ and vice versa \cite{Evenbly2018}. 
}
\label{fig:NTUg}
\end{figure}

\section{Algorithms}
\label{sec:algorithms}

The algorithms considered in this paper are summarized in figures \ref{fig:2site}, \ref{fig:2site2}, \ref{fig:env}, \ref{fig:NTUenv}, and \ref{fig:NTUg}. Figures \ref{fig:2site}(b,c) show the most basic singular value decomposition update (SVDU) in a schematic form. After a two-site Trotter gate is applied to a pair of nearest neighbor iPEPS tensors, $A$ and $B$, the resulting network in Fig. \ref{fig:2site}(b) is SV-decomposed into a pair of new tensors $A'$ and $B'$. The dimension of their common bond index is truncated to the original $D$ by keeping only the $D$ largest singular values. Numerical cost of this scheme is $\propto D^9$. Its equivalent but more efficient version is shown in Fig. \ref{fig:2site2}. The cost is cut down to $\propto D^5$ by reduction to smaller matrices $R_{A,B}$ before the SVD truncation. The truncation yields new reduced matrices $M_{A,B}$ that are fused with fixed isometries $Q_{A,B}$ into updated iPEPS tensors $A'$ and $B'$.

The SVDU minimizes the Frobenius norm of the difference between diagrams in Figs. \ref{fig:2site2}(b) and (d). For this norm all directions in the $(Dr)^2$-dimensional space are equally important. Thus, though formally cheap, the SVDU does not make optimal use of the available bond dimension $D$ which is wasted to preserve accuracy in all directions including those that are not important from the perspective of the infinite tensor environment of the two sites. Even zero modes, that are not important at all, instead of being truncated are preserved as accurately as the dominant directions. On the positive side, the SVDU is inverse free.

A step beyond SVDU, whose cost is still $\propto D^5$, is the simple update (SU) \cite{Orus_review_14}. In this scheme the iPEPS ansatz in Fig. \ref{fig:2site}(a) is generalized by inserting its bonds with diagonal bond tensors $\lambda_i$, where $i$ is numbering four inequivalent bonds on the checkerboard lattice. The Frobenius norm is replaced by a metric
\be 
g_{\rm SU} = 1_d \otimes 1_d \otimes \prod_j \lambda_j,
\ee 
where $j$ runs over the six bonds stemming out from the considered pair of NN sites. These bonds are the nearest tensor environment providing the nontrivial metric tensor that assigns different weights to different directions. SU can afford the same bond dimension as SVDU but, in principle, can make better use of it. A potential caveat is inversion of the bond tensors: $\lambda_i\to\lambda_i^{-1}$ that has to be done after every gate. 

In this paper we advocate a step beyond the SU where a cluster of nearest neighbor (NN) tensors, shown in Fig. \ref{fig:NTUenv}(b), is the environment providing the metric. This NN cluster can be contracted exactly, as outlined in Fig. \ref{fig:NTUenv}, to yield metric $g$ that is Hermitian and non-negative within machine precision. The cost of optimal contraction is $\propto D^{8}$ but, as it involves only matrix multiplication, can be fully parallelized. The key advantage of the metric in Fig. \ref{fig:NTUenv}(e) over the local SVDU/SU are the two NN bonds, parallel to the considered one, that connect the left and right side of the environment. They are essential to prevent virtual loop entanglement from being build into the iPEPS and parasite its bond dimension. We call the scheme a neighborhood tensor update (NTU) to distinguish it from a full tensor update (FTU), where the infinite environment in Fig. \ref{fig:NTUenv}(a) provides the metric tensor. 

This infinite environment is the same as in the popular full update (FU) scheme \cite{Orus_review_14}. FU and FTU differ in the way the iPEPS tensors are decomposed into isometries $Q_{A,B}$ and reduced tensors/matrices $R_{A,B}$. In this paper both schemes serve mainly as a benchmark. Their infinite environment takes into account long range correlations but calculation of the metric tensor $g$ requires an expensive corner transfer matrix renormalization group (CTMRG) \cite{Orus_review_14} whose approximate character makes it difficult to keep the metric tensor Hermitian and non-negative. In NTU the CTMRG is used only for calculation of expectation values which can be done less frequently and may require less precision than the Trotter gates. 

With metric tensor $g$ matrices $M_A$ and $M_B$ are optimized in order to minimize the norm squared of the difference between the two diagrams in Fig. \ref{fig:NTUg}(a), where $R_A R_B^T$ is the exact (untruncated) product in Fig. \ref{fig:2site2}(b). The error is measured with respect to the metric in Fig. \ref{fig:NTUenv}(d,e):
\be 
\varepsilon=\left[M_AM_B^T-R_AR_B^T\right]^\dag ~ g ~ \left[M_AM_B^T-R_AR_B^T\right].
\ee 
For a fixed $M_B$ it becomes a quadratic form in $M_A$:
\be 
\varepsilon = M_A^\dag g_A M_A - M_A^\dag J_A - J_A^\dag M_A + \varepsilon_A,
\ee 
where $g_A$, $J_A$, and $\varepsilon_A$ depend on the fixed $M_B$, see Fig. \ref{fig:NTUg}(b) and (c). The matrix is optimized as
\be 
M_A={\rm pinv}\left(g_A\right)J_A,
\ee 
where tolerance of the pseudo-inverse can be dynamically adjusted to minimize $\varepsilon$. Thanks to the exactness of $g$ in NTU, the optimal tolerance is usually close to machine precision. This optimization of $M_A$ is followed by a similar optimization of $M_B$. The optimizations are repeated in a loop,
\be 
\rightarrow M_A \rightarrow M_B \rightarrow,
\ee 
until convergence of $\varepsilon$. Except for SVD of small matrices, $R_AR_B^T$ and $M_AM_B^T$, NTU is fully parallelizable. 

\begin{figure}[t!]
\vspace{-0cm}
\includegraphics[width=1.0\columnwidth,clip=true]{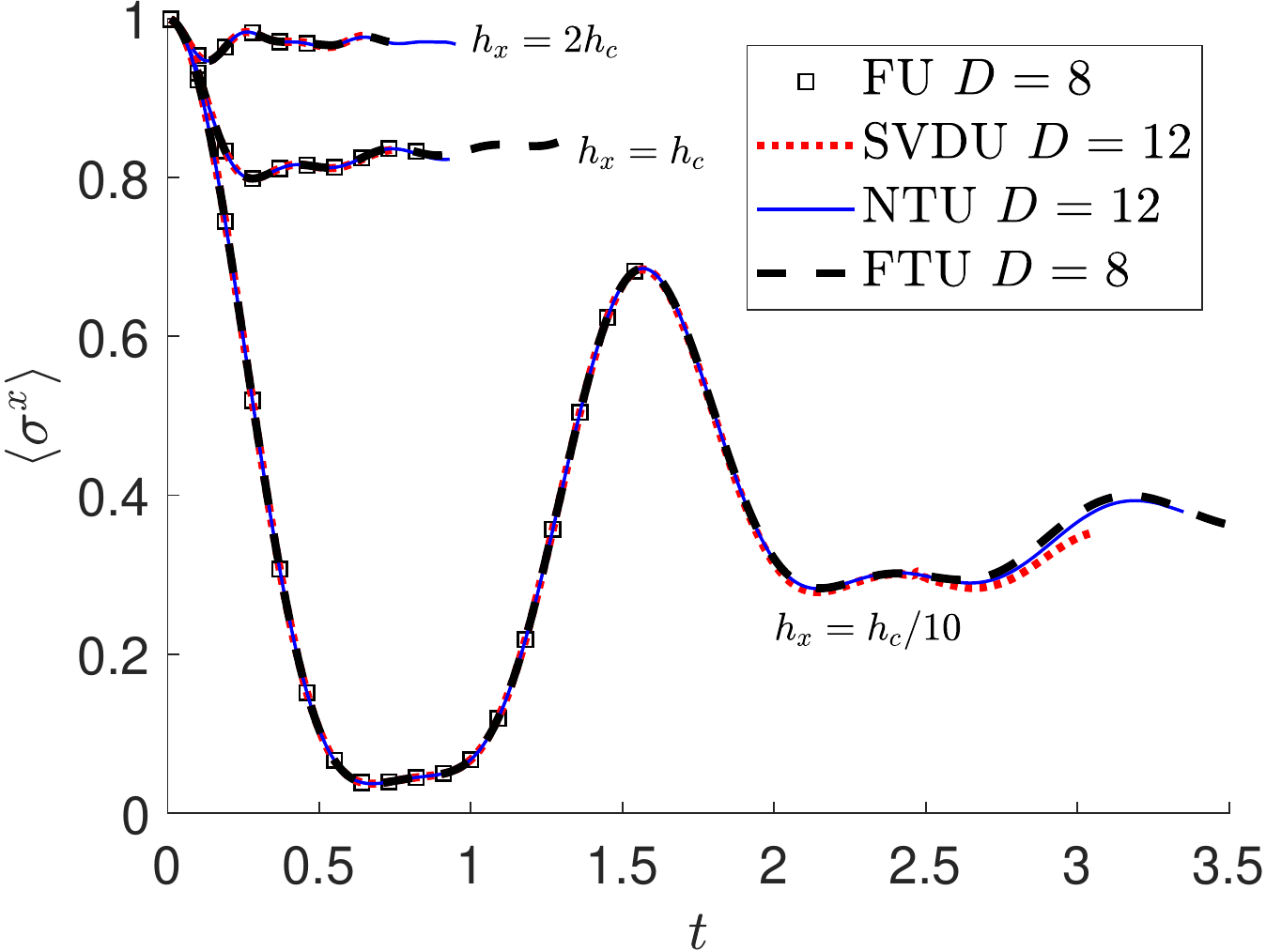}
\vspace{-0cm}
\caption{
{\bf Sudden quench.} 
Unitary evolution of the transverse magnetization $\langle \sigma^x \rangle$ after a sudden quench from a fully polarized state. We show three bunches of curves corresponding to evolution with (from top to bottom) $h_x=2h_c,h_c,h_c/10$. Each curve is terminated when the energy per site deviates by $0.01$ from the initial value. The squares are data from FU simulations\cite{CzarnikDziarmagaCorboz}. They extend up to the time where they appear converged in $D$ for $D=8$ (for $h_x=2h_c,h_c$) or up to $t=\pi/2$ where they were terminated (for $h_x=h_c/10$). Here we use the same time step, $dt=0.01$, as for the FU\cite{CzarnikDziarmagaCorboz}, the same second order Suzuki-Trotter scheme and environmental bond dimension: $\chi=4D$. 
}
\label{fig:sudden}
\end{figure}

\begin{figure}[t!]
\vspace{-0cm}
\includegraphics[width=0.99\columnwidth,clip=true]{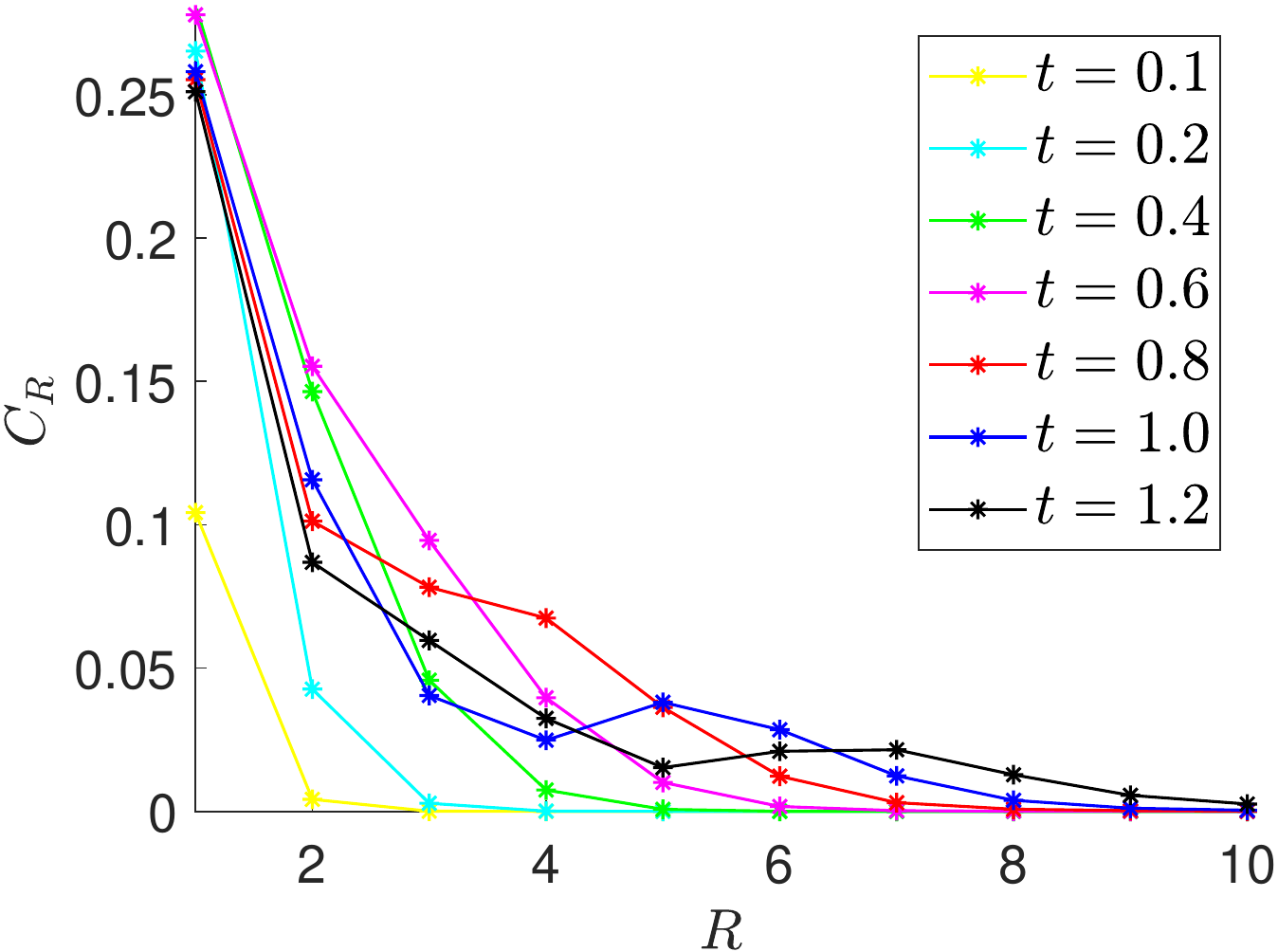}
\vspace{-0cm}
\caption{
{\bf Sudden quench.} 
Connected correlation function, 
$
C_R=
\langle\sigma^z_n\sigma^z_{n+R}\rangle-
\langle\sigma^z_n\rangle\langle\sigma^z_{n+R}\rangle
$ 
at different times after the sudden quench to $h_x=h_c$.
}
\label{fig:sudden_corr}
\end{figure}

\section{Unitary evolution after a sudden quench}
\label{sec:sudden}

To begin with we consider a sudden quench in the transverse field quantum Ising model on an infinite square lattice:
\be 
H_{\rm QI} =
-\sum_{\langle j,j'\rangle} \sigma^z_j \sigma^z_{j'} - \sum_j h_x \sigma^x_j.
\label{HI}
\ee
At zero temperature the model has a ferromagnetic phase with non-zero spontaneous magnetization $\langle \sigma^z \rangle$ for magnitude of the transverse field, $|h_x|$, below a quantum critical point located at $h_c=3.04438(2)$\cite{Deng_QIshc_02}. 

Here we simulate unitary evolution after a sudden quench at time $t=0$ from infinite transverse field down to a finite $h_x$. After $t=0$ the fully polarized ground state of the initial Hamiltonian is evolved by the final Hamiltonian with $h_x=2h_c,h_c,h_c/10$. The same quenches were simulated with FU \cite{CzarnikDziarmagaCorboz} and neural quantum states \cite{ANN_Markus&Markus}. Our present results obtained with SVDU, NTU, and FTU are shown in Fig. \ref{fig:sudden}. As a benchmark we also show the FU results with $D=8$ up to times where they appear converged with this bond dimension. 

The evolution with the weakest $h_x=h_c/10$ remains weakly entangled for a long time and can be extended to long simulation times by any iPEPS method. This is not surprising given that for $h_x=0$, when the Hamiltonian is classical, exact evolution can be represented with mere $D=2$. All the considered simulation schemes reproduce the exact $D=2$ evolution for $h_x=0$. The quenches to $h_x=2h_c,h_c$ are more challenging as they create a lot of entanglement. 

We show SVDU, NTU, and FTU results with, respectively, $D=12,12,8$. These bond dimensions require similar simulation time as FU with $D=8$. All simulations, except FU, are terminated when the energy per site deviates by more than $0.01$ from its initial value. For all three $h_x$ NTU provides longer evolution time than SVDU, as expected. Relation between FTU and other schemes is not quite systematic because, unlike the other schemes, FTU often ends by a sudden crash that makes its evolution time somewhat erratic. Nevertheless, in the most challenging quench to the critical point, $h_x=h_c$, FTU outperforms the other schemes. This is expected as in this case correlation range developed after the quench is the longest, see Fig. \ref{fig:sudden_corr}.

The sudden quench benchmark encourages applications of NTU to other time-dependent problems. The first in row is the Kibble-Zurek finite rate quench that was simulated by NTU in Ref. \onlinecite{schmitt2021quantum} where its results were corroborated by neural networks \cite{ANN_Markus&Markus} and matrix product states.

\begin{figure}[t!]
\vspace{-0cm}
\includegraphics[width=0.9999\columnwidth,clip=true]{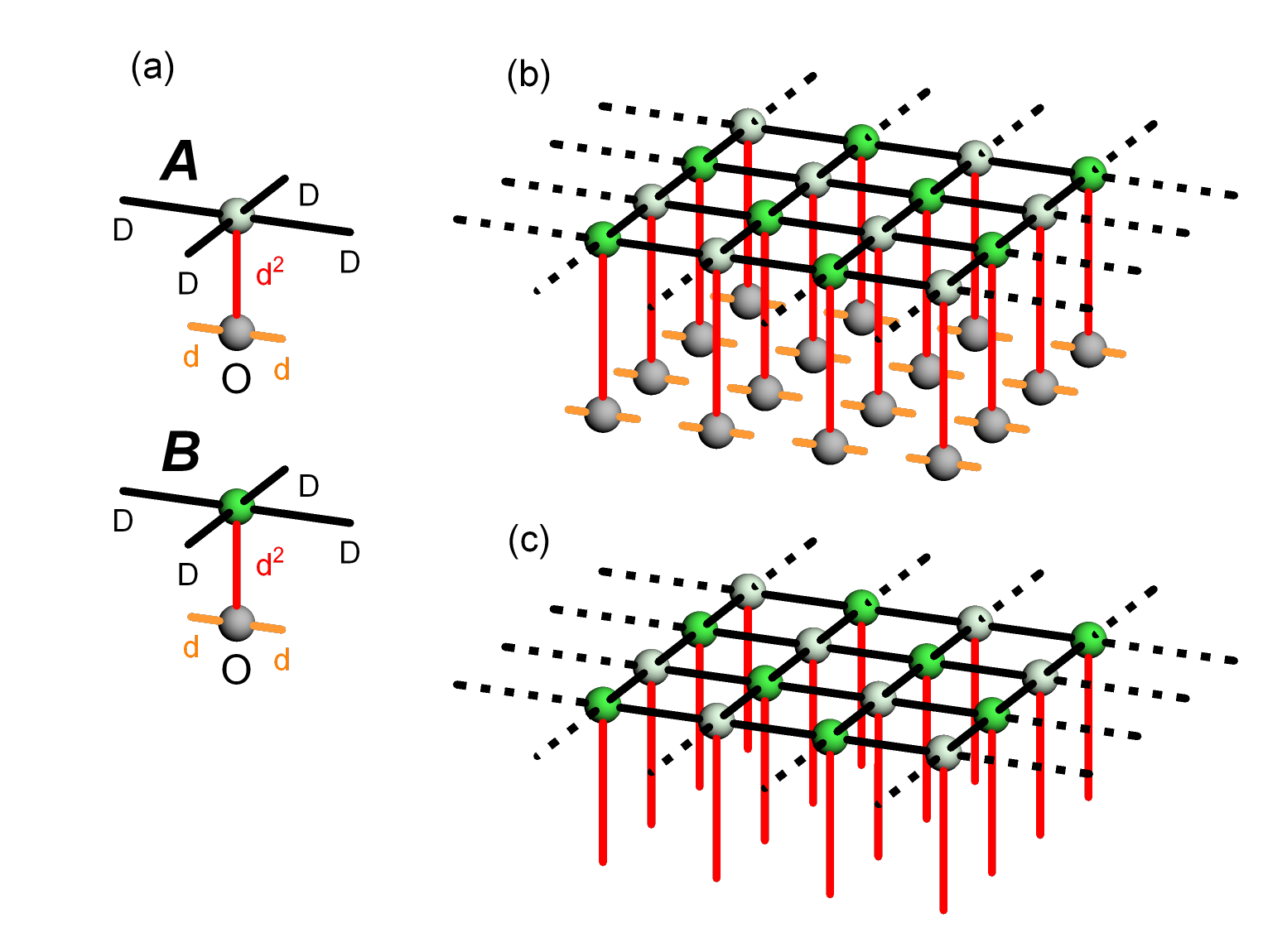}
\vspace{-0cm}
\caption{
In (a)
at every lattice site real rank-5 iPEPS tensor, either $A$ or $B$, has four bond indices of dimension $D$ -- to be contracted with similar tensors on its four NN sites -- and an index of dimension $d^2$ contracting it with tensor $O$.
Tensor $O$ has additional two indices, bra and ket, each of dimension $d$ equal to the dimension of the Hilbert space at every lattice site. $O^a$, where $a$ is the index of dimension $d^2$, is a basis of Hermitian operators. For spin-$1/2$ with $d=2$ we choose $O^1=\sigma^x,O^2=\sigma^y,O^3=\sigma^z,O^4=1$.
In (b)
when contracted through their bond indices the rank-6 tensors in (a) make an iPEPO - a tensor network representation of a Hermitian operator between bra and ket indices.
In actual computations we are dealing only with the top part of (b) which is an iPEPS $\tilde\rho$ made of real tensors.
Its physical indices have dimension $d^2$.
}
\label{fig:Hgauge}
\end{figure}

\section{Simulation of thermal states}
\label{sec:thermal}

In a series of tautologies a thermal state, $\rho(\beta)\equiv e^{-\beta H}$, can be written as $\rho(\beta)=\rho(\beta/2)\rho(\beta/2)$, where
\be 
\rho(\beta/2)=e^{-\beta H/4}\rho(0/2)\left(e^{-\beta H/4}\right)^\dag.
\ee 
We represent this $\rho(\beta/2)$ as an iPEPO and the evolution operator is a product of small time steps,
\be 
e^{-\beta H/4}=\prod_{i=1}^N e^{-d\beta H/4}.
\ee
A small time step for the iPEPO is
\be 
\rho[(\beta+d\beta)/2]=
e^{-d\beta H/4}
\rho(\beta/2)
\left(e^{-d\beta H/4}\right)^\dag,
\ee 
where $e^{-d\beta H/4}$ is approximated by a Suzuki-Trotter decomposition into a product of Trotter gates. $\rho[(\beta+d\beta)/2]$ would be manifestly Hermitian if it were not necessary to truncate the bond dimension after each Trotter gate. 

In order to preserve the Hermitian symmetry in Fig. \ref{fig:Hgauge} we introduce a manifestly Hermitian parametrization of the iPEPO. In effect, the iPEPO $\rho$ is represented by an iPEPS $\tilde\rho$ made of real tensors. In addition to manifestly preserving the symmetry, that may improve numerical stability, this real parametrization should speed up floating number computations by a factor of $4$.  

In the next section we test the algorithm in the 2D quantum Ising model, where the non-trivial nearest-neighbor 2-site Trotter gate is
\be 
e^{d\beta \sigma^z_j \sigma^z_{j'}/4} \propto
1_j1_{j'} + \tanh(d\beta/4) \sigma^z_j\sigma^z_{j'} \equiv G.
\ee 
Under its action the basis operators $O^a_jO^b_{j'}$, defined in Fig. \ref{fig:Hgauge}, transform as
\bea
G~O^aO^b~G^\dag = \sum_{a'b'} G^{ab}_{a'b'} O^{a'} O^{b'}.
\eea
Therefore, the action of the gate $G$ on the iPEPO, $G \rho G^\dag$, is equivalent to contracting the iPEPS $\tilde\rho$ with the tensor $G^{ab}_{a'b'}$. The upper indices of the latter, $a$ and $b$, are contracted with the physical indices of the iPEPS on sites $j$ and $j'$, respectively, as shown in Fig. \ref{fig:2site}(b). The SVDU, NTU, and FTU algorithms follow as in section \ref{sec:algorithms}.

\begin{figure}[t!]
\vspace{-0cm}
\includegraphics[width=0.9999\columnwidth,clip=true]{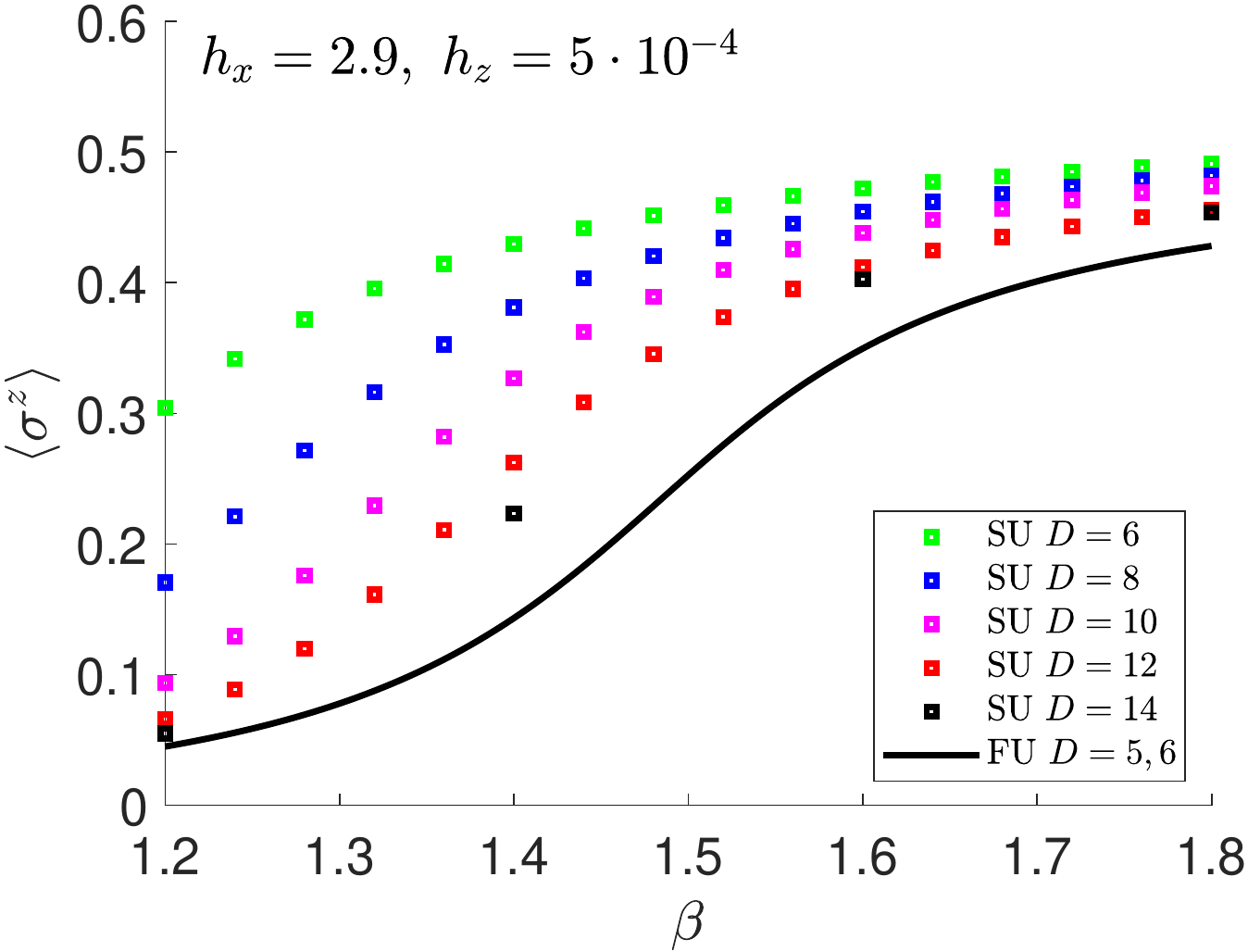}
\vspace{-0cm}
\caption{
{\bf Thermal states by SU and FU. }
Comparison of thermal states obtained by SU and FU evolution of a thermal state purification for $h_x=2.9$ and $h_z = 5\cdot 10^{-4}$. With increasing $D$ the SU magnetization curve moves slowly towards the converged FU magnetization with $D=5,6$ but even for the largest $D=14$ it is still far from it. All data in this figure come from Ref. \onlinecite{CzarnikDziarmagaCorboz}.
}
\label{fig:ISTHold}
\end{figure}

\begin{figure}[t!]
\vspace{-0cm}
\includegraphics[width=0.9999\columnwidth,clip=true]{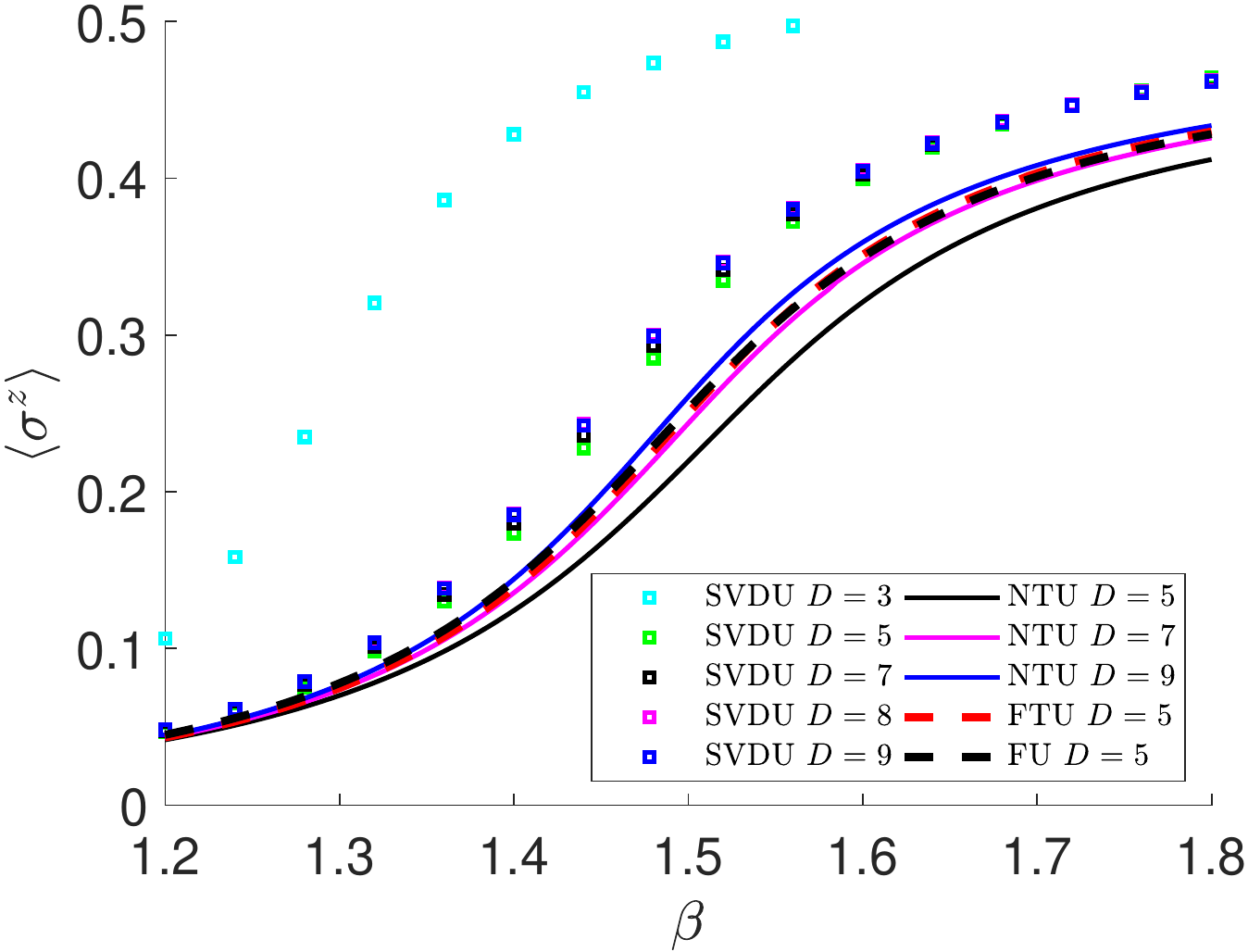}
\includegraphics[width=0.9999\columnwidth,clip=true]{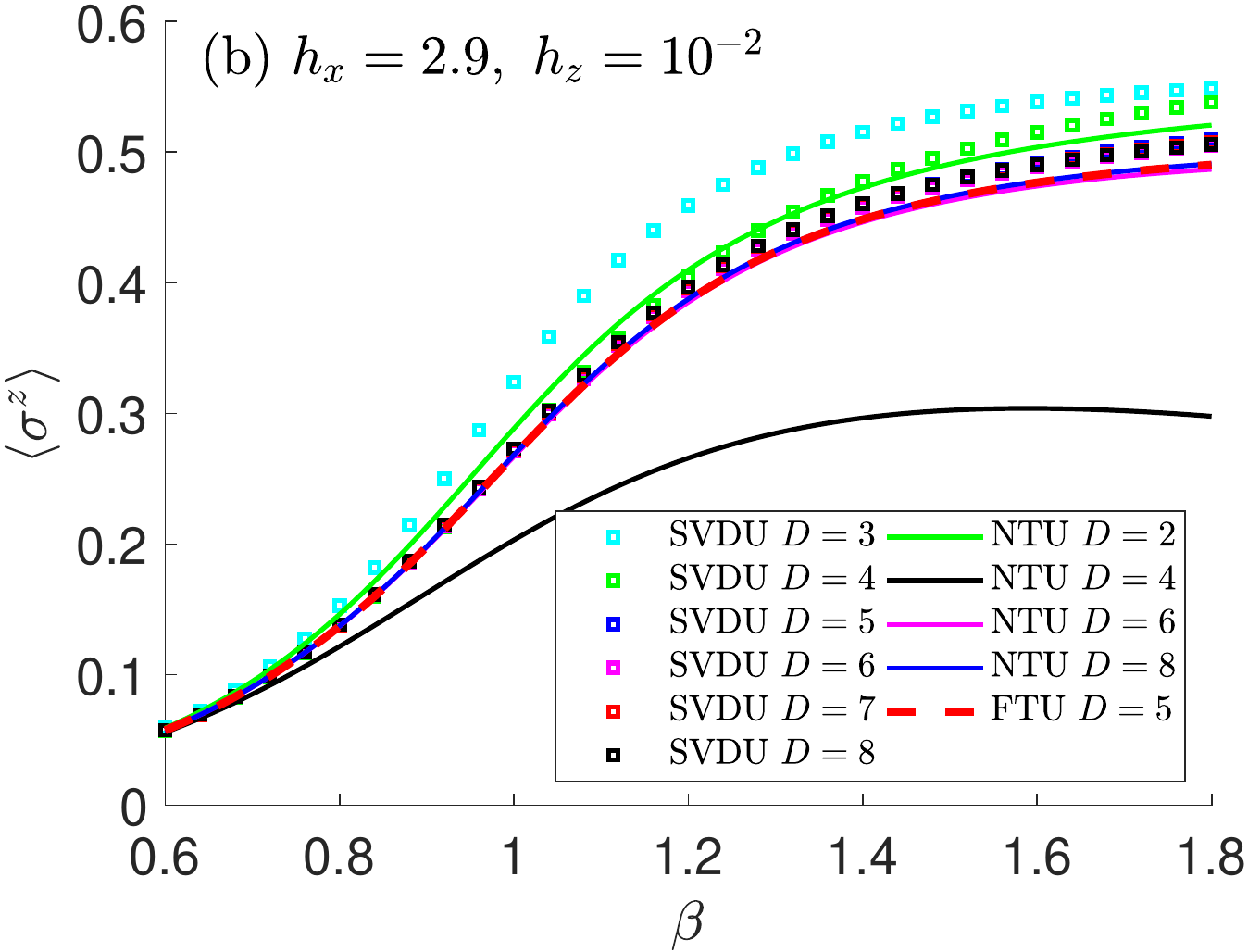}
\vspace{-0cm}
\caption{
{\bf Thermal states by SVDU, NTU, and FTU.} 
In (a) thermal states obtained with FTU and SVDU for $h_x=2.9$, which is very close to the quantum critical $h_c=3.04438(2)$\cite{Deng_QIshc_02}, with a weak bias $h_z = 5\cdot 10^{-4}$. The FTU and FU appear converged for bond dimension $D=5$ and they serve as a benchmark for SVDU and NTU. SVDU gets close to the benchmark as its $D$ grows from $3$ to $6$ but for $D=7,8,9$ slightly drifts away from it. NTU gets close to FTU/FU for $D=7..9$.
In (b) the same as in (a) but with a stronger bias $h_z = 10^{-2}$. Again, FTU appears converged for $D=5$ and serves as a benchmark for SVDU and NTU. SVDU converges towards the benchmark as $D$ is increased to $6$. For $D=7,8$ it drifts up from it but much less than for the weaker bias in panel (a). NTU is converged to FTU for $D\geq6$.
}
\label{fig:ISTHnew29}
\end{figure}

\begin{figure}[t!]
\vspace{-0cm}
\includegraphics[width=0.9999\columnwidth,clip=true]{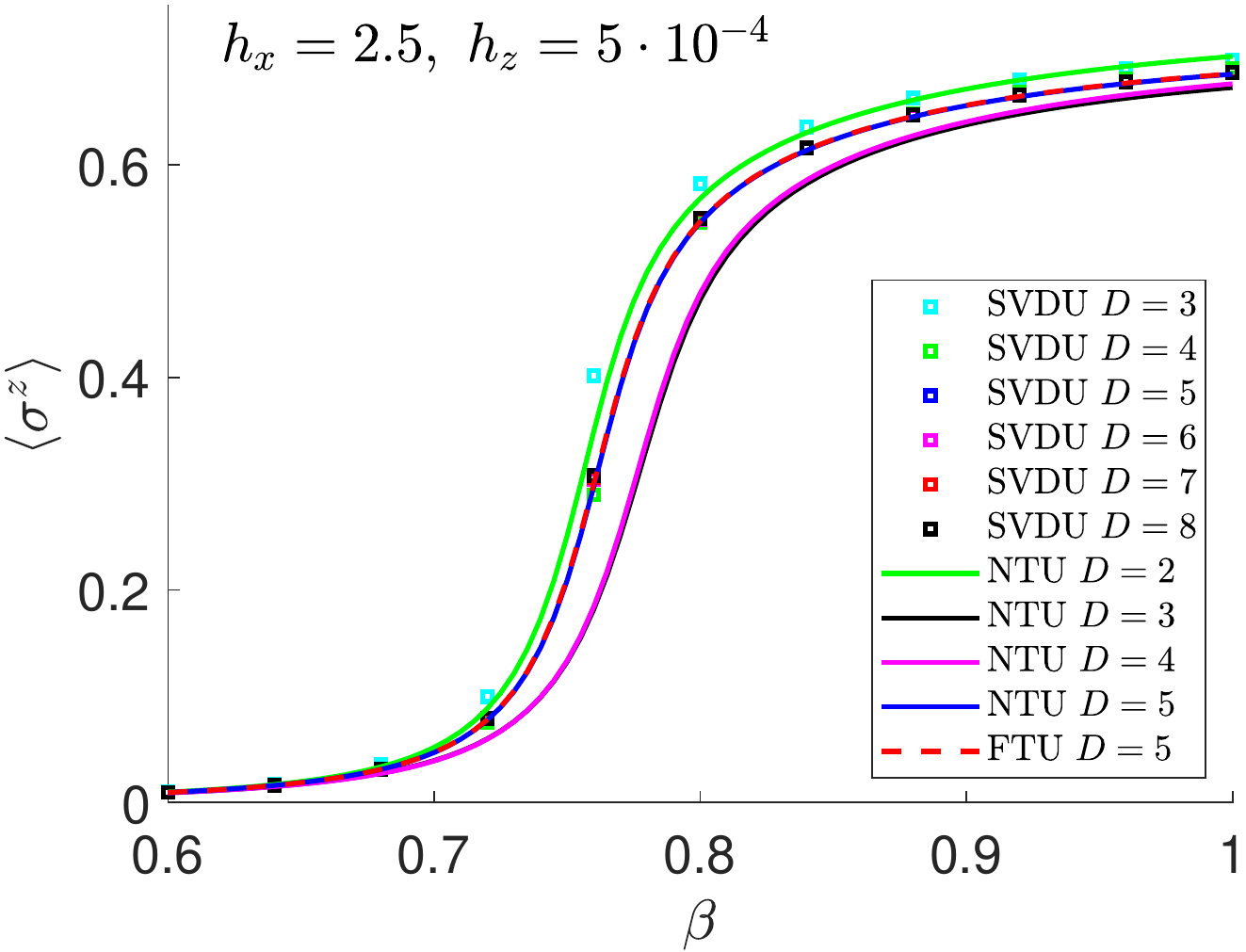}
\vspace{-0cm}
\caption{
{\bf Thermal states by SVDU, NTU, and FTU.} 
Thermal states for $h_x=2.5$, which is further away from the quantum critical $h_c=3.04438(2)$\cite{Deng_QIshc_02}, with a weak bias $h_z = 5\cdot 10^{-4}$. FTU appears converged for bond dimension $D=5$ which serves as a benchmark for SVDU and NTU. SVDU gets close to the benchmark as its $D$ grows up to $6$. For $D=7,8$ it slightly drifts up from it but not in an appreciable way. NTU is converged for $D\geq5$.
}
\label{fig:ISTHnew25}
\end{figure}

\section{2D quantum Ising model at finite temperature}
\label{sec:thermalIsing}

The Hamiltonian of the quantum Ising model with a longitudinal bias on an infinite square lattice is
\bea 
H&=&
H_{\rm QI} -\sum_j h_z \sigma^z .
\label{HIbias}
\eea
Here $H_{\rm QI}$ is the transverse field quantum Ising model \eqref{HI} and $h_z$ is a longitudinal field providing
a tiny symmetry-breaking bias that allows for smooth evolution across a finite-temperature phase transition
by converting it into a smooth crossover. For zero longitudinal field and $h_x < h_c$ the model has a second order phase transition at a finite temperature, $T_c(h_x)$, belonging to the 2D classical Ising universality class. For $h_x=0$ it becomes the 2D classical Ising model with $T_c(0)=2/\ln(1+\sqrt{2})\approx 2.27$. 

In all simulations in this section we use $d\beta=0.0025$ and the second order Suzuki-Trotter decomposition. The data are converged in the environmental bond dimension which is set at $\chi=40$. Finally, in all FTU simulations we begin with a short SVDU evolution stage up to $\beta=0.1$. This avoids dealing with zero modes which arise when the bond dimension is too big \cite{CzarnikSS}.

First we consider $h_x=2.9$ --- which is very close to $h_c=3.04438(2)$\cite{Deng_QIshc_02} --- where the critical temperature is estimated as $T_c(2.9)=0.6085(8)$\cite{Hesselmann_TIsingQMC_16}. Due to strong quantum fluctuations this is almost four times less than the Onsager's $T_c(0)$. We generate thermal states across this transition with a bias field $h_z=5\cdot10^{-4}$ which is one of the weakest biases considered in Ref. \onlinecite{CzarnikDziarmagaCorboz} where the same states were obtained with SU and FU schemes. The SU data\cite{CzarnikDziarmagaCorboz} in Fig. \ref{fig:ISTHold} show that under these extreme conditions SU is not able to converge to the converged FU results (with $D=5,6$) even for the largest considered bond dimension $D=14$. Pushing the simulations beyond $D=14$ becomes more costly than the more accurate FU and thus becomes impractical\cite{CzarnikDziarmagaCorboz}. 

Figure \ref{fig:ISTHnew29}(a) shows new FTU results which are converged for $D=5$ similarly as the old FU. Quite remarkably, as the bond dimension in SVDU is increased from $D=3$ up to a mere $D=6$, which is still very cheap for this local update, the results get closer to the converged FTU results than the SU ones with $D=14$. For $D=6$ a maximal correlation length $\xi\approx15$ is achieved at $\beta=1.44\dots1.48$ which is more than might have been expected from a local update. However, this record is a warning sign that anticipates the following decline in accuracy as the bond dimension is increased further beyond $D=6$. The decline is most visible for $\beta=1.44\dots1.48$ where the record long correlations make the local update method the most problematic. The same Fig. \ref{fig:ISTHnew29}(a) shows results from NTU as they slowly converge for $D=7,...,9$. The converged NTU curve slightly differs from the FTU one but much less than the SVU results.

In order to see how SVDU and NTU perform under less severe conditions, in Fig. \ref{fig:ISTHnew29}(b) we show results for the same $h_x=2.9$ but with a stronger bias $h_z=10^{-2}$. Again, $D=5$ is enough to converge FTU. With $D$ growing from $3$ to $6$ the SVDU gets much closer to the converged FTU benchmark than for the weaker bias. Beyond $D=6$ some decline in accuracy is observed but it is much less significant than for the weaker bias. The better convergence can be explained by a much shorter correlation length which peaks at $\xi\approx4$ near $\beta=1$. The same correlation length explains why the NTU magnetization curves with $D\geq6$ coincide with the FTU one.

In order to see if the correlation length is the sole factor determining quality of the SVDU/NTU convergence, we move away from the quantum critical point down to $h_x=2.5$ and consider again the weaker bias $h_z=5\cdot10^{-4}$. The critical temperature is $T_c(2.5)=1.2737(6)$\cite{Hesselmann_TIsingQMC_16} which is a little more than half of $T_c(0)$ indicating that quantum fluctuations are much less influential than for $h_x=2.9$ but still significant. The results are shown in Fig. \ref{fig:ISTHnew25}. Again, FTU is converged for $D=5$ and SVDU is the closest to the FTU benchmark for $D=6$ and slightly drifts up for $D=7,8$ but this time the difference between SVDU and FTU is negligible: SVDU with $D=5,6,7,8$ are practically converged to the benchmark though they have some scatter. The correlation length calculated at $\beta=0.76$ is $\xi\approx 22$, i.e., the longest of the three examples. In spite of this it does not prevent convergence of either SVDU or NTU: NTU is converged already for $D=5$. Therefore, it is not the correlation length alone that matters but the quantum nature of the correlations.

\begin{table}[t!]
\begin{tabular}{|l|c|c|c|}
\hline
method & $D$ & $T_c$ & $1/\tilde\beta\delta$ \\ 
\hline
SU\cite{CzarnikDziarmagaCorboz} & $12$ & $0.704(11)$ & $0.85(11)$ \\
\hline
  NTU  & $5$  &  $0.5858(28)$ & $0.586(7)$ \\ 
  NTU  & $6$  &  $0.5995(38)$ & $0.606(11)$ \\   
  NTU  & $7$  &  $0.6021(21)$ & $0.611(6)$ \\   
  NTU  & $8$  &  $0.6084(42)$ & $0.611(13)$ \\ 
  NTU  & $9$  &  $0.6089(40)$ & $0.618(14)$ \\ 
\hline   
FU\cite{CzarnikDziarmagaCorboz} & $5$ & $0.6100(7)$ & $0.571(3)$ \\
\hline
QMC\cite{Hesselmann_TIsingQMC_16} & - & $0.6085(8)$ & - \\
\hline
exact & - & - & $8/15 \approx 0.533$\\
\hline
\end{tabular}
\caption{
Comparison of $T_c$ and $1/\tilde\beta\delta$ obtained with NTU for $h_x=2.9$ and the bias in the range: $0.0005 \le h_z \le 0.01$. 
For comparison we also list the SU and FU results \cite{CzarnikDziarmagaCorboz}.
The quantum Monte Carlo estimate \cite{Hesselmann_TIsingQMC_16} is shown as a bechmark.
In brackets we show 95\% confidence intervals. 
}
\label{tab:29}
\end{table}

\begin{table}[t!]
\begin{tabular}{|l|c|c|c|c|}
\hline
method  & $D$ &  $T_c$ & $1/\tilde\beta\delta$ \\ 
\hline
  NTU  & $2$ &  $1.2820(10)$ & $0.576(13)$ \\  
  NTU  & $3$ &  $1.2430(20)$ & $0.570(9)$ \\  
  NTU  & $4$ &  $1.2450(20)$ & $0.573(11)$ \\  
  NTU  & $5$ &  $1.2740(20)$ & $0.578(14)$ \\  
  NTU  & $6$ &  $1.2740(15)$ & $0.579(13)$ \\  
\hline
FU\cite{CzarnikDziarmagaCorboz} & $5$ & $1.2745(7)$ & $0.549(4)$  \\
\hline
QMC\cite{Hesselmann_TIsingQMC_16} & - & $1.2737(6)$ & - \\
\hline
exact &   -   &      -      & $8/15 \approx 0.533$\\
\hline
\end{tabular}
\caption{
Comparison of $T_c$ and $1/\tilde\beta\delta$ obtained with NTU for $h_x=2.5$ and the bias in the range: 
$0.00035 \le h_z \le 0.0056$. 
For comparison we also list the FU result \cite{CzarnikDziarmagaCorboz}.
The quantum Monte Carlo estimate \cite{Hesselmann_TIsingQMC_16} is shown as a benchmark.
In brackets we show 95\% confidence intervals. 
}
\label{tab:25}
\end{table}

\begin{figure}[t!]
\vspace{-0cm}
\includegraphics[width=0.9999\columnwidth,clip=true]{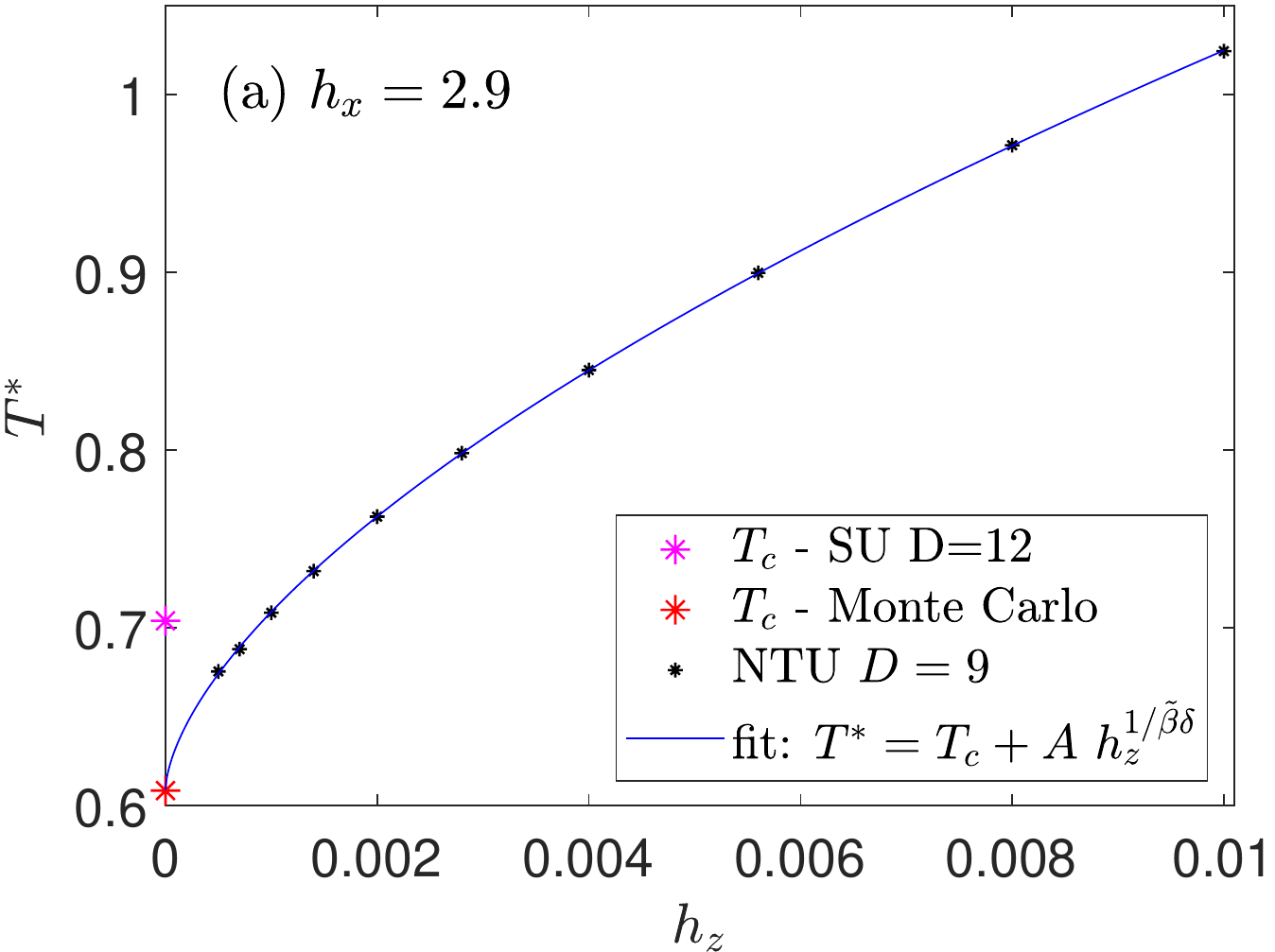}
\includegraphics[width=0.9999\columnwidth,clip=true]{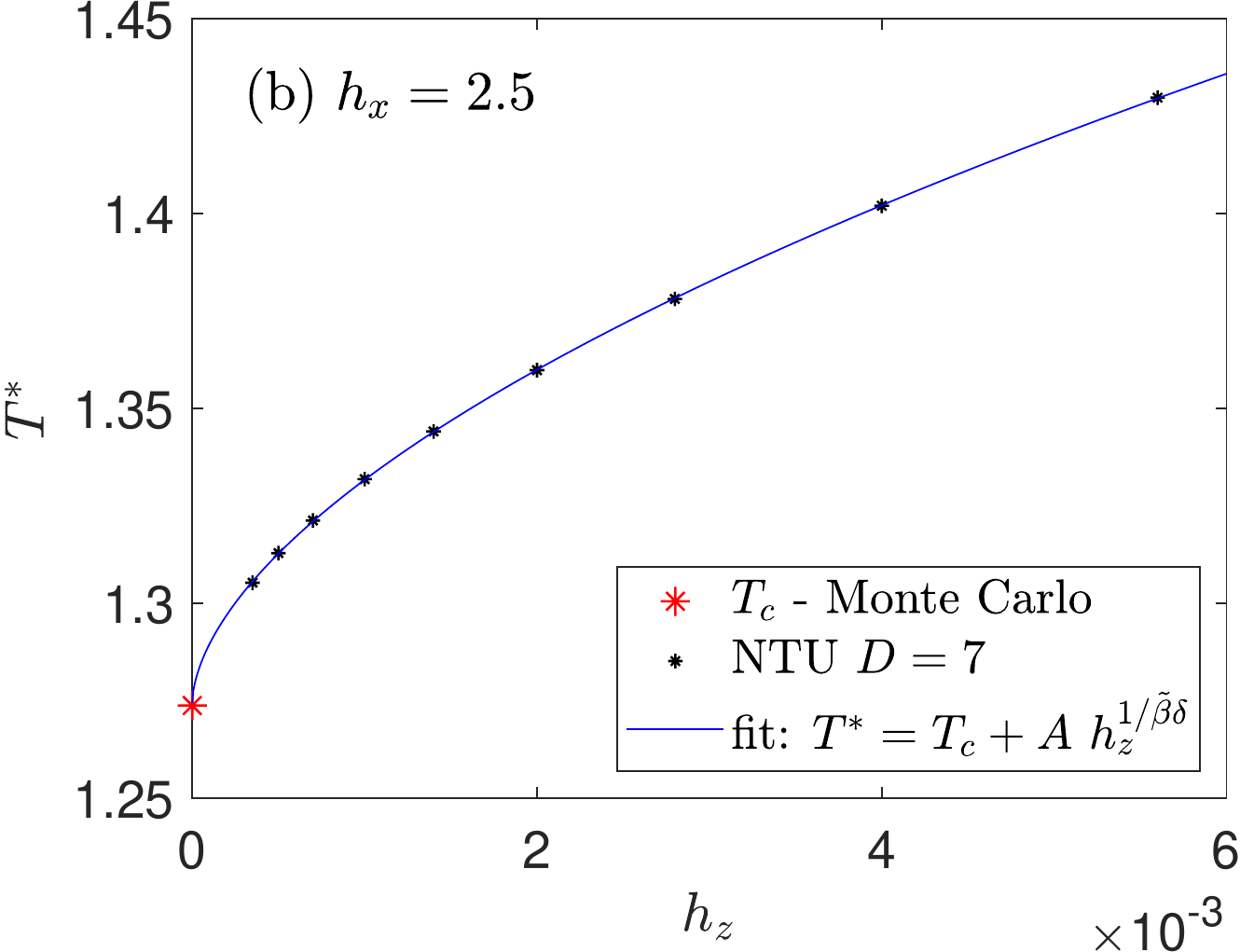}
\vspace{-0cm}
\caption{
{\bf Critical temperature from NTU. } 
Pseudo-critical temperature $T^*$ --- obtained as the temperature where magnetization $\langle\sigma^z\rangle$ is the steepest as a function of $\beta$ --- in function of bias $h_z$ is fitted with the power law in \eqref{T*} for $h_x=2.9$ in (a) and $h_x=2.5$ in (b). Corresponding quantum Monte Carlo estimates for $T_c$ are marked with red stars. 
In (a) SU estimate \cite{CzarnikDziarmagaCorboz} is shown as a magenta star. $T_c$'s estimated for different bond dimensions and their error bars are listed in tables \ref{tab:29} and \ref{tab:25}.
}
\label{fig:fits}
\end{figure}

The convergence of the NTU results encourages us to attempt estimation of critical temperature $T_c(h_x)$ from magnetization curves --- $\langle\sigma^z\rangle$ in function of $\beta$ --- obtained for different $h_z$, see Ref. \cite{CzarnikDziarmagaCorboz} for more details of the procedure. For each $h_z$ we find a pseudo-critical temperature, $T^*(h_z)$, where the slope of the magnetization in function of $\beta$ is the steepest. Then we make a fit:
\be 
T^*(h_z)=T_c+A~ h_z^{1/\tilde\beta\delta},
\label{T*}
\ee 
where $\tilde\beta,\delta$ are critical exponents. Treating $T_c,A$ and $1/\tilde\beta\delta$ as fitting parameters we obtain estimates of critical temperatures $T_c$ for $h_x=2.9$ and $h_x=2.5$ that are listed in tables \ref{tab:29} and \ref{tab:25}, respectively. The best fits \eqref{T*} are shown in Fig. \ref{fig:fits}. For both values of transverse field NTU yields estimates of $T_c$ that are consistent with those from FU\cite{CzarnikDziarmagaCorboz} and quantum Monte Carlo \cite{Hesselmann_TIsingQMC_16} although their convergence requires higher $D$ than FU. The error bars are wider than for FU and the exponent, $1/\tilde\beta\delta$, is more overestimated. 

\section{Conclusion}
\label{sec:conclusion}

We considered three evolution algorithms that can be ordered according to their increasing size of tensor environment that is taken into account when optimizing tensors: SVDU, NTU, and FTU. In general, the increasing size translates to faster convergence with bond dimension $D$. On this scale the traditional SU sits between SVDU and NTU while FTU is a variant of FU:
$$ 
{\rm SVDU} < {\rm SU} < {\rm NTU} < {\rm FTU} \approx {\rm FU}.
$$
The increasing environment correlates with increasing numerical cost. However, in the latter respect NTU is in practice not much more expensive than SU. Although formally its cost of calculating the neighborhood environment scales like $D^8$, as compared to the leading cost of $D^5$ for SU, the $D^8$ is a fully parallelizable tensor contraction while the $D^5$ is a non-parallelizable SVD. When compared with FTU/FU, on the other side, NTU convergence with $D$ is in general slower but, thanks to the numericaly exact environment, it offers more stability/efficiency for higher $D$ that allow to compensate for the limitations of the small environment. Therefore, for many applications NTU may be an attractive alternative for SU and FU alike.

\acknowledgements
%
I would like to thank Aritra Sinha and Piotr Czarnik for comments on the manuscript.
This research was supported in part by the National Science Centre (NCN), Poland 
under projects 2019/35/B/ST3/01028 (JD).
%
\appendix
\bibliography{ref.bib} 


\end{document}